\documentclass[PP]{AEA}

\usepackage{natbib}

    \usepackage[dvipsnames]{xcolor}
	\usepackage{amsfonts}
	\usepackage{amssymb}
	\usepackage{amsmath}
	
	\usepackage{enumerate}
    \usepackage{accents}
	
	\usepackage{hyperref}
	\hypersetup{
    colorlinks=true,
    linkcolor=Red,
    filecolor=magenta,      
    urlcolor=blue,
    citecolor = Blue
}
 \usepackage{cleveref}

    \usepackage{rotating}
	\usepackage[framemethod=tikz]{mdframed}
	\usepackage{todonotes}
	\presetkeys{todonotes}{color=black!5,inline}{} 

    \usepackage{tcolorbox}
    \newtcbox{\feedback}{nobeforeafter,colframe=black,colback=white,boxrule=0.5pt,arc=2pt,
      boxsep=0pt,left=2pt,right=2pt,top=2pt,bottom=2pt,tcbox raise base}
      
    \newtheorem{prop}{Proposition}

\usepackage{array}
\newcolumntype{L}[1]{>{\raggedright\let\newline\\\arraybackslash}m{#1}}
\newcolumntype{C}[1]{>{\centering\let\newline\\\arraybackslash\hspace{0pt}}m{#1}}
\newcolumntype{R}[1]{>{\raggedleft\let\newline\\\arraybackslash\hspace{0pt}}m{#1}}

\usepackage{ragged2e}
\newlength\ubwidth

\newcommand{\onevec}{\mathbf{1}}

	\newcommand{\parens}[1]{\left(#1\right)}

	\newcommand{\reals}{\mathbb{R}}

	\newcommand{\betahat}{\ensuremath{\hat{\beta}}}

\newcommand{\normnot}[2]{\mathcal{N}\parens{#1,\,#2}}

\newcommand{\twovec}[2]{\parens{\begin{array}{c} #1 \\ #2 \end{array}}}

\newcommand{\betahatpost}{\betahat_{post}}

\newcommand{\betahatpre}{\betahat_{pre}}

\newcommand{\betapost}{\beta_{post}}

\newcommand{\deltapost}{\delta_{post}}
\newcommand{\deltapre}{\delta_{pre}}

\newcommand{\taupost}{\tau_{post}}

\newcommand{\betapre}{\beta_{pre}}

\newcommand{\ubar}[1]{\underaccent{\bar}{#1}}

\draftSpacing{1}

\usepackage{xr}
\begin{document}

\title{(Empirical) Bayes Approaches to Parallel Trends}
\shortTitle{(Empirical) Bayes Parallel Trends}
\author{Soonwoo Kwon and Jonathan Roth\thanks{\hspace{-6.5pt} Kwon: \hspace{-1pt}Brown \hspace{-1pt}University, \hspace{-1pt}\href{mailto:soonwoo\_kwon@brown.edu}{soonwoo\_kwon@brown.edu};
Roth: Brown University, \href{mailto:jonathanroth@brown.edu}{jonathanroth@brown.edu}. We are grateful to Dmitry Arkhangelsky, Kirill Borusyak, Peter Hull, and Ashesh Rambachan for comments.}}\pubMonth{Month}
\date{\today}
\pubYear{Year}
\pubVolume{Vol}
\pubIssue{Issue}
\maketitle














Researchers employing a difference-in-differences (DiD) design are often unsure about the validity of the parallel trends assumption. It is common to test for ``pre-trends'', yet such tests may be underpowered, and relying on them leads to statistical issues from pre-testing \citep{roth_pretest_2022}. Recent work by \citet{manski_how_2017} and \citet[][RR]{rambachan_more_2023} has made progress on obtaining more credible inference when parallel trends may be violated by adopting a partial identification approach. In RR, for example, the researcher places bounds that restrict the possible values of the post-treatment violations of parallel trends $\deltapost$ given the identified pre-treatment violations $\deltapre$. The identified set for the treatment effect then corresponds to the worst-case bounds for $\deltapost$ given the observed $\deltapre$. 

We instead consider a Bayesian approach where the researcher imposes a prior on the violations of parallel trends $\delta$. The researcher then updates their posterior about $\deltapost$ given the observed estimate of $\deltapre$. This allows them to form posterior means and credible sets (CSs) for the treatment effect $\taupost$. The Bayesian approach allows the researcher to impose ex ante information about what violations of parallel trends may look like, and thus to potentially obtain more informative results than the partial identification approach using worst-case bounds. It also allows one to form point estimates in addition to confidence sets. For settings with many pre-treatment periods, we also consider empirical Bayes (EB) approaches, where the ``prior'' for the violations of parallel trends is calibrated using the pre-trends. 

For more general work on Bayesian approaches in settings with partial identification, see \citet{moon_bayesian_2012, giacomini_robust_2021}. For a related EB approach to parallel trends, see \citet{leavitt_beyond_2020}. 
\section{Set-up}
Following RR, we consider a setting where the researcher observes a vector of event-study estimates $\betahat = (\betahatpre', \betahatpost')' \in \reals^{\ubar{T} + \bar{T}}$ corresponding to $\ubar{T}$ pre-treatment and $\bar{T}$ post-treatment periods. Motivated by asymptotics based on the central limit theorem, we suppose that $\betahat$ is normally distributed with known variance, $\betahat \sim \normnot{\beta}{\Sigma_{\betahat}}$, where 
\begin{equation}
            \beta = \twovec{0}{\taupost} + \twovec{\deltapre}{\deltapost} . \label{eqn: beta decomposed to tau and delta}
        \end{equation}
The vector $\tau$ corresponds to the treatment effect in each period (assumed to be zero prior to treatment, $\tau_{pre} = 0$), while $\delta$ corresponds to a vector of biases (e.g. violations of parallel trends). RR consider restrictions that impose that $\delta \in \Delta$. This enables partial identification of $\tau$, with bounds corresponding to worst-case assumptions on the element $\delta \in \Delta$. In this paper, we alternatively consider Bayesian inference where the researcher places a prior on $\delta$, as well as Empirical Bayes approaches where the prior on $\deltapost$ is calibrated using $\deltapre$. 

\section{Fully Bayesian Approach}
We impose a prior $\pi_{\tau, \delta}(\cdot)$  over $\tau, \delta$.\footnote{For notational convenience, in this section we write $\tau$ for a vector of the form $(0,\taupost')'$.} From Bayes' rule, we have that 
$$p( \tau, \delta \mid \betahat ) \, \propto \, \ell(\betahat \mid \delta + \tau) \cdot \pi_{\tau, \delta}(\tau, \delta),$$
where $\ell( \betahat \mid \beta)$ denotes the normal likelihood of observing $\betahat$ given $\betahat \mid \beta \sim \normnot{\beta}{\Sigma_{\hat{\beta}}}$. Consequently, 
\begin{align*}
p(\tau \mid \betahat) =& \int p( \tau, \delta \mid \betahat) \, d\delta  \\
\propto & \int \ell(\betahat \mid \delta + \tau) \cdot \pi_{\tau,\delta}(\tau,\delta) \, d\delta 
\end{align*}

\noindent Thus, given a prior $\pi_{\tau,\delta}$ it is straightforward to compute the posterior $p(\tau \mid \betahat)$.

In what follows, we will primarily consider the case where the researcher has an uninformative
prior on $\tau \mid \delta$, so that $\pi_{\tau \mid \delta}(\tau \mid \delta)
\,\propto\, 1 $, in which case
$$p(\tau \mid \betahat) \, \propto \, \int \ell(\betahat \mid \delta + \tau)
\cdot \pi_\delta(\delta) \, d\delta.$$

The following result characterizes the posterior mean for $\taupost$ when the prior is uninformative.

\begin{prop} \label{prop: posterior mean}
If the prior for $\tau$ is uninformative (i.e. $\pi_{\tau \mid \delta} \,\propto \, 1$), then
\begin{align*}
&E[\taupost \mid \betahat] = \\ &E[ \betapost \mid \betahat ] - \underbrace{ E[ E[\deltapost \mid \deltapre = \betapre] \mid \betahat ] }_{= E[\deltapost \mid \betahat]}    
\end{align*}
\end{prop}

\noindent Proposition \ref{prop: posterior mean} shows that the posterior mean for $\taupost$ is simply the difference between the posterior for $\betapost$ and the posterior for $\deltapost$. It shows further that the posterior for $\deltapost$ can be written as an iterated expectation, where the inner expectation is based on the conditional prior of $\deltapost$ given $\deltapre$, and the outer expectation is over the posterior for $\betapre \mid \betahat$.

It is worth noting that the expression for $E[\taupost \mid \betahat]$ derived in Proposition \ref{prop: posterior mean} depends on the \emph{conditional} prior on the post-treatment bias $\deltapost$ given the pre-trend $\deltapre$, regardless of the precision of the estimates $\betahat$. This reflects the well-known fact that in partially identified settings, the prior matters \emph{even asymptotically}. Researchers adopting this approach must therefore be careful to choose a prior that reflects economic information about the possible violations of parallel trends.

\paragraph{Example: Gaussian Prior} Suppose we have a Gaussian prior for $\delta$, $\delta \sim \normnot{\mu_\delta}{V_\delta}$ and impose the uninformative prior for $\tau$. A straightforward calculation using Bayes' rule shows that the posterior for $\taupost$ is also Gaussian. To derive the posterior mean, note that the formula for the conditional mean of a Gaussian vector implies that 
$$E[ \deltapost \mid \deltapre ] =  \mu_{\deltapost} + \Gamma_V' (\deltapre - \mu_{\deltapre}) ,$$

\noindent for $\Gamma_V = V_{\deltapre}^{-1} V_{\deltapre,\deltapost}$. Applying Proposition \ref{prop: posterior mean},
$$E[ \taupost \mid \betahat ] = \beta^*_{post} - \mu_{\deltapost} - \Gamma_V' (\betapre^* - \mu_{\deltapre}) ,$$
\noindent where $\beta^* = E[\beta \mid \betahat]$ is the posterior mean for $\beta$. 

One can further show that the posterior mean for $\betapre$ is
\begin{align*}
\betapre^* = (\Sigma_{\betahatpre}^{-1} + V_{pre}^{-1} )^{-1} (\Sigma_{\betahatpre}^{-1} \betahatpre + V_{pre}^{-1} \mu_{\deltapre}),  
\end{align*}
\noindent which ``shrinks'' the point-estimate $\betahatpre$ towards the prior mean $\mu_{\deltapre}$. Likewise, the posterior mean for $\betapost$ is
\begin{align*}
\betapost^* &= \betahatpost - \Gamma_\Sigma' (\betahatpre - \betapre^*)     
\end{align*}

\noindent for $\Gamma_\Sigma = \Sigma_{\betahatpre}^{-1} \Sigma_{\betahatpre,\betahatpost}$. See the Online Appendix for detailed calculations, a formula for the posterior variance of $\taupost$, and an extension to the case with an uninformative Gaussian prior on $\taupost$. 

\paragraph{Empirical Illustration}

\citet[][BZ]{benzarti_who_2019} study the impacts of a reduction in the value-added tax on restaurants in France. They run a non-staggered DiD design comparing profits for restaurants to those of firms in other industries not affected by the tax change. The key concern with this approach is that there might be idiosyncratic economic factors affecting the profits of restaurants that do not affect other industries, which would lead to violations of parallel trends. We calibrate our prior on these violations using \citet{mcgahan_persistence_1999}, who estimate an $AR(1)$ process for the industry-level component of firm profits (see the Online Appendix for detail). We take their estimates of the $AR(1)$ parameters and assume that the $AR(1)$ innovations are from a mean-zero Gaussian, which implies a Gaussian prior for the violations of parallel trends. The figure below shows the original OLS estimates and confidence intervals (CIs) from BZ, as well as posterior means and 95\% CSs from our Bayesian approach.

\begin{center}
\includegraphics[width = 0.99 \linewidth]{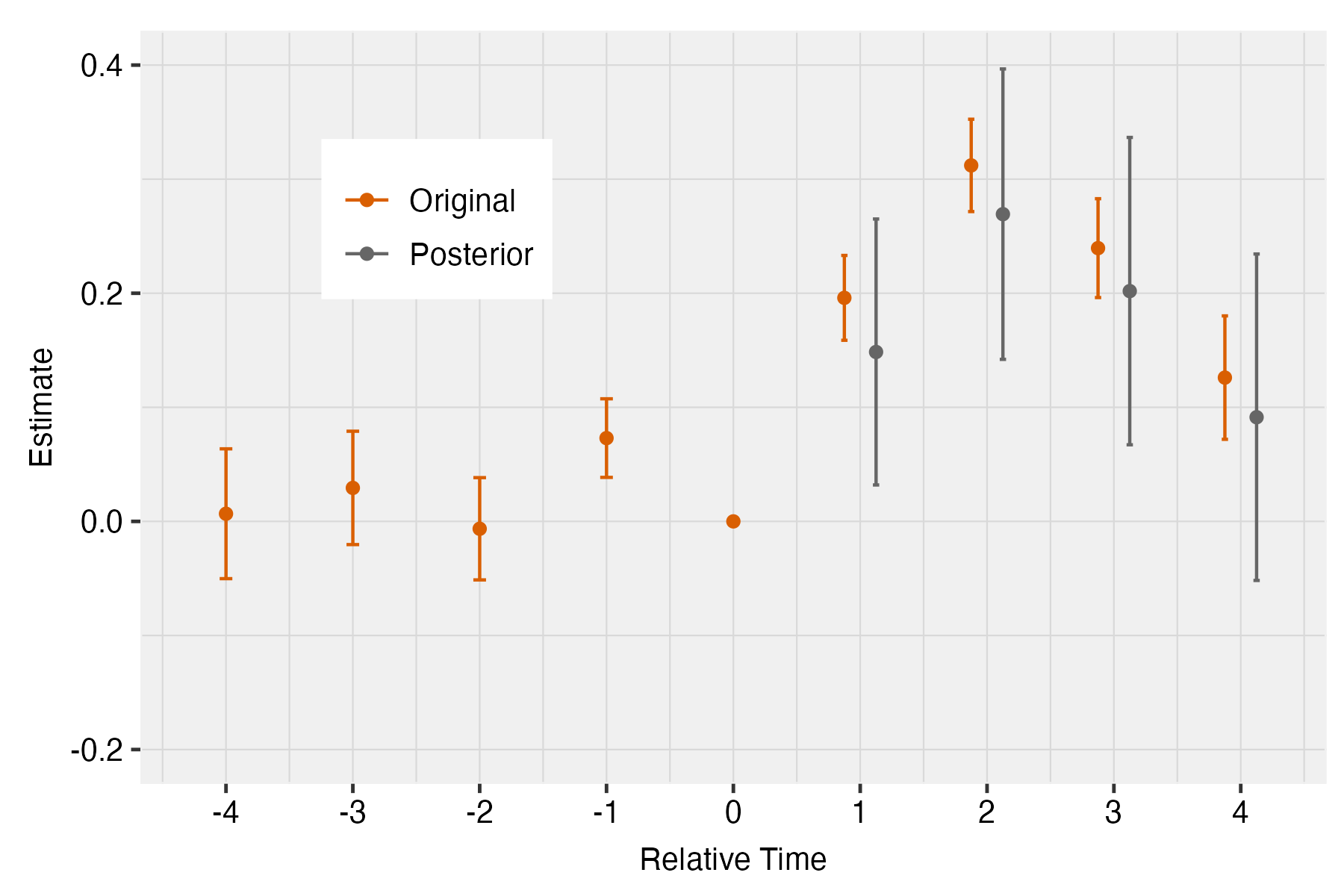}
\end{center}

The posterior CSs are wider than the OLS CIs, since the OLS CIs assume that parallel trends holds exactly, whereas our prior puts positive weight on violations of parallel trends. Nevertheless, the CSs are informative, excluding zero in 3 out of 4 post-treatment periods. The posterior means are also somewhat closer to zero than the OLS estimates. This is because $\deltapre$ and $\deltapost$ are correlated under the imposed prior; thus, the primarily positive estimates for $\betahatpre$ lead to a posterior that the post-treatment bias $\deltapost$ is positive. 

\section{Empirical Bayes Approaches}

We saw in the previous section that the conditional prior $\deltapost \mid \deltapre$ matters regardless of the precision of the event-study estimates $\betahat$. This conditional prior governs how violations of parallel trends evolve over time. In settings where we have many pre-treatment periods, and we think that the violations of parallel trends come from a \emph{stationary} process, it might be attractive to learn the time-series dependence of violations of parallel trends from the pre-trends. This motivates an EB approach where the parameters of the time series process for violations of parallel trends are learned from the pre-trends, and posterior estimates are then calculated based on the prior implied by the estimated parameters. 

As a simple illustration, suppose that violations of parallel trends across consecutive periods are governed by $w_t := \delta_t - \delta_{t-1} \overset{iid}{\sim} \normnot{\mu}{\sigma^2}$. In a simple non-staggered DiD, this corresponds to the case where the idiosyncratic factors differentially affecting the treated group follow a Gaussian random walk with drift. Let $w_{pre} = (w_{-\ubar{T}+1},...,w_{0})'$ collect the pre-treatment values of $w_t$. Analogously define the vector $\hat{w}_{pre}$ to collect the estimate of $w_{pre}$ using $\betahatpre$ instead of $\deltapre$. Then we have that $\hat{w}_{pre} \sim \normnot{ \mu \cdot \onevec }{ \Sigma_w + \sigma^2 I }$, where $\onevec$ is the vector of ones, and $\Sigma_w = M \Sigma_{\betahatpre} M'$ for $M$ the matrix such that $\hat{w}_{pre} = M \betahatpre$. The parameters $\mu$ and $\sigma^2$ can thus be estimated via maximum likelihood, which will be consistent (under mild regularity conditions on $\Sigma_w$) as the number of pre-treatment periods grows large, $\ubar{T} \to \infty$. Since the assumption that $w_t \overset{iid}{\sim} \normnot{\mu}{\sigma^2}$ implies a normal prior for $\delta$,\footnote{We adopt the common normalization that $\delta_0 =0$, which allows us to infer the distribution of $\delta$ from $w$.} it is straightforward to calculate the posterior for $\taupost$ using the prior implied by the estimates $\hat\mu, \hat\sigma$.  

One caveat to this approach is that the consistency of the estimates for the prior depends on the number of pre-treatment periods $\ubar{T}$ being large. In practice, the number of pre-treatment periods may be moderate---e.g., in our empirical application below, it is 9---in which case estimates based on this approach must be interpreted with some caution. We note that an alternative to the EB approach when the number of periods is moderate is to consider a hierarchical Bayes model, where one imposes a hyper-prior on the parameters $\mu,\sigma^2$ and then updates their prior based on the observed estimate of $\betahatpre$. This approach retains validity even when $\hat\mu,\hat\sigma$ are only imprecisely estimated, but of course requires the researcher to specify a prior on the hyper-parameters.

\paragraph{Empirical Illustration}

\citet{lovenheim_long-run_2019} study how being exposed to laws that increase the power of teachers' unions as a student impacts earnings in adulthood. They use a two-way fixed effects event-study specification exploiting the differential timing of the passage of these laws.\footnote{A recent literature surveyed in \citet{roth_whats_2023} has shown that such specifications may be difficult to interpret under treatment effect heterogeneity; one could re-do the analysis here with the event-study from one of the estimators developed for this case.} The concern with the parallel trends assumption is that states passing these laws may have different secular trends in labor market outcomes. To address this, we suppose that $w_t \overset{iid}{\sim} \normnot{\mu}{\sigma^2}$, and estimate the parameters $\mu,\sigma$ using maximum likelihood based on the pre-trends. 

\includegraphics[width = 0.99 \linewidth]{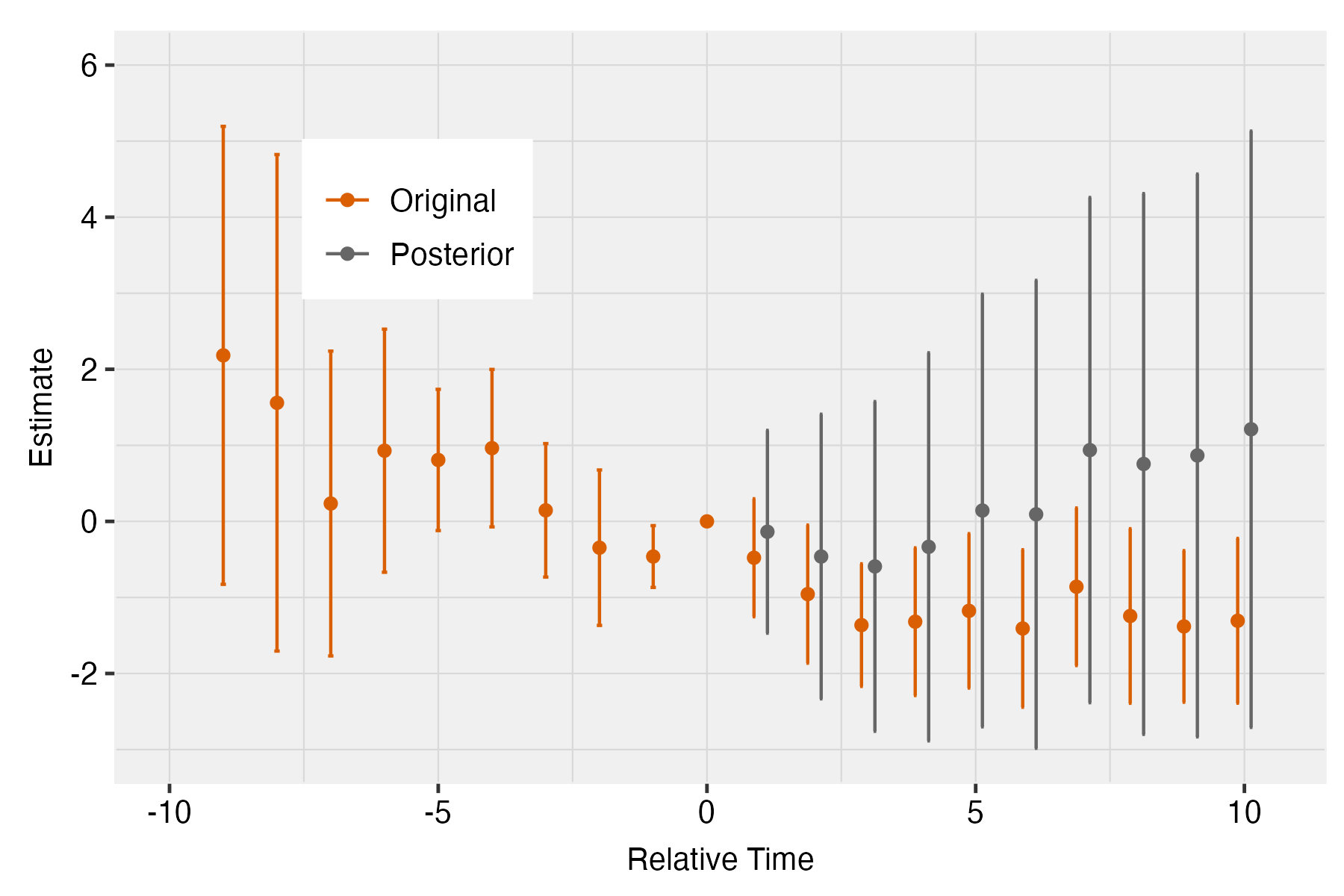}

Using female employment (in p.p.) as the outcome, we estimate $\hat\mu = -0.24, \hat\sigma = 0.61$, indicating a downward-sloping pre-trend and some variance around it. Because of the prior that the violation of parallel trends is downward sloping, the posterior means for the treatment effects are substantially closer to zero than the OLS estimates (see figure above). The CSs are also substantially wider than the OLS CIs, especially in later post-treatment periods, owing to uncertainty about the violations of parallel trends.

\bibliographystyle{aea}
\bibliography{Bibliography}

\appendix

\end{document}